# Using Steered Molecular Dynamic Tension for Assessing Quality of Computational Protein Structure Models

Lyman Monroe[1] and Daisuke Kihara[1,2,3,*]

[1] Department of Biological Sciences, Purdue University, West Lafayette, IN 47907, USA
[2] Department of Computer Science, Purdue University, West Lafayette, IN 47907, USA
[3] Purdue University Center for Cancer Research, West Lafayette, IN, 47907, USA

[*] Correspondence Author: dkihara@purdue.edu

**Abstract**

The native structures of proteins, except for notable exceptions of intrinsically disordered proteins, in general take their most stable conformation in the physiological condition to maintain their structural framework so that their biological function can be properly carried out. Experimentally, the stability of a protein can be measured by several means, among which the pulling experiment using the atomic force microscope (AFM) stands as a unique method. AFM directly measures the resistance from unfolding, which can be quantified from the observed force-extension profile. It has been shown that key features observed in an AFM pulling experiment can be well reproduced by computational molecular dynamics simulations. Here, we applied computational pulling for estimating the accuracy of computational protein structure models under the hypothesis that the structural stability would positively correlated with the accuracy, i.e. the closeness to the native, of a model. We used in total 4,929 structure models for 24 target proteins from the Critical Assessment of Techniques of Structure Prediction (CASP) and investigated if the magnitude of the break force, i.e. the force required to rearrange the model's structure, from the force profile was sufficient information for selecting near-native models. We found that near-native models can be successfully selected by examining their break forces suggesting that high break force indeed indicates high stability of models. On the other hand, there were also near-native models that had relatively low peak forces. The mechanisms of the stability exhibited by the break forces were explored and discussed.

**Introduction**

It is generally understood that the native structure of a protein adopts the most thermodynamically stable fold in the protein's conformation space[1] except for some notable examples including of intrinsic disordered proteins and prions[2]. An implication of this postulation, which is called the thermodynamic hypothesis, is that the thermodynamically stable structures of proteins are dictated by the amino acid sequence under physiological conditions, leading to the development of methods for protein structure prediction from amino acid sequence. In structure prediction, a method often generates tens to thousands of possible conformations for a single protein sequence. This required the development of a sub-field of structure prediction, which is known as protein quality assessment (QA)[3]. QA is aimed at predicting the accuracy of computational structure models or rank models in a model pool based on their expected accuracy. Strategies of QA include application of statistical and physical potential functions[4-6], evaluating consistency to stereochemistry of models to known structures[7,8], consistency of models with predicted local structures and alignments to known structures[9,10], consensus with alternative models[11,12], and machine learning approaches that combine various structural features of models[13-16].

Here, we evaluated computational protein structure models by directly measuring their stability in virtual pulling experiment of protein models. We used Steered Molecular Dynamics (SMD), which mimics pulling experiments with Atomic Force Microscopy (AFM). AFM conventionally scans a sample by a probe and can measure the force between the probe and the sample surface or can be applied to image the sample shape[17]. With regard to proteins, AFM has been used to measure the binding force between a protein and a ligand[18], to characterize enzyme activity[19], to study unfolding pathways[20-22] and mechanical stability of proteins[23]. Early applications of SMD used to mimic AFM on biological systems were performed by the Schulten group to study the unbinding of the avidin-biotin complex[24] and the unfolding pathway of titin IG domains[25], which showed good agreement with cotemporary experiments using AFM[26]. Recently, simulated AFM was applied to investigate unfolding of the cold shock protein B (Csp)[27] and the Src SH3 domain[28].

In this study we hypothesized that, under MD-simulated stress, models with a high structural similarity to their native conformation would be more resistant to unfolding than models with more inaccurate folds. Using a set of computational models submitted to the Critical Assessment of Techniques for protein Structure Prediction (CASP) [29-31], a community-wide experiment of protein structure prediction, we investigated if models that are close to the native would be selected by examining the peak of the forces measured in the simulated pulling experiment. We found that near-native models can be successfully selected by examining the magnitude of their break forces, that is the force required to rearrange the structure, suggesting that high break forces indeed indicates high stability of models. On the other hand, there were also models that were similarly close to the native yet had relatively lower break forces. By comparing these two groups, it was determined that among near-native models, the force required for structural rearrangement was dependent on three primary factors; bonded interactions, electrostatic interactions, and solvent interactions.

**Materials and Methods**

**Data Set**

We used the submitted prediction models for CASP10, 11, and 12, which are available at the CASP website[32-34]. The files were available to download at https://predictioncenter.org/download_area/CASP10/SUMMARY_TABLES/ for CASP10 files, and at the corresponding locations for targets from CASP 11 and 12. From the CASP data, target sets were selected for use if the sets included models with GDT-TS (global distance test total score)[35] below 50.0 and above 85.0. GDT-TS computes the average of the percentage of C$\alpha$ atoms in a model that are modelled within 1.0, 2.0, 4.0, and 8.0 Å and thus ranges from 0 to 100. After filtering by GDT-TS scores, we then removed any target set which were either part of a larger complex or were associated with lipid molecules, as the stabilizing effects of either a binding partner or lipid environment would not be present in the simulation. Finally, any models with disulfide bonds were removed, as the molecular dynamics would not break these disulfide bonds and would rapidly become unphysical. Filtering at this point left us with 24 target sets with a total of 4,929 models to test. The average number of models for a target was 205.38. The average minimum, maximum, and average GDT_TS of models for a target were, 15.84, 92.17, and 75.48, respectively. The distribution of models across GDT_TS ranges for each target is shown in Table 1. GDT-TS. values of models were taken from the tables that associated with the model files from the CASP website.

For each model, we produced two additional variant models with identical backbone structure but different sequences. These different sequence models consisted of one in which all residues were converted to alanine, and one where the sequence was reversed. This was done by aligning an all alanine and a reversed sequence with the original model's sequence, and then producing a structure using MODELLER[36]. The purpose of these additional models was to introduce a sequence variable to the sequence-structure relationship. Converting the sequence of each protein to all alanine removed the specific sequence characteristics and unique sidechain interactions that contributed to the energetic profile of the fold. Reversing the sequence was for the same purpose but retaining the same set of atoms and total mass as the original model.

**Pulling a structure model using Molecular Dynamics**

All molecular dynamics (MD) simulations were carried out using NAMD[37]. The MD protocol consisted of four phases; first, a psf file that contains atom bonds and angle information of a target protein structure was generated using VMD's autopsf function. and the models were then solvated with TRIP3P water molecules using the VMD autopsf plug-in, solvate[38]. Second, the models were minimized using a step size of 1 femto seconds (fs) for 100 pico seconds (ps) to remove atomic clashes. Subsequently, the models were equilibrated with the temperature increasing from 10 to 300 Kelvin over 30 ps and then further equilibrated for 0.5 nano seconds (ns). Finally, the structures were simulated under a steered MD (SMD) framework for 1 ns.

In the production run, an SMD atom with no mass or charge was placed in the simulation with a spring with a spring constant of 7 kcal/mol/Å$^2$ connecting the dummy atom with the C-terminal alpha-carbon. This dummy atom was them moved at a constant velocity of 0.001 Å per fs along the vector connecting the N-terminal alpha-carbon to the C-terminal alpha carbon. Meanwhile, the N-terminal alpha-carbon was fixed in place. This caused the dummy atom to pull the protein until it unfolded and eventually linearized. By calculating the tension on the simulated spring, we determined the force required to pull apart and rearrange domains of the protein throughout the simulation. Figure 1 shows an example of a force profile of a protein model. From the profile, the force required for rearrangement was taken as the first substantial peak in a trajectory, highlighted with a red circle at II in Figure 1. Along the trajectory, the structure of the model is shown below the profile panel. Stage I shows the initial structure before the simulation begins. Stage II shows the structure at high tension just before rearrangement. At this point, the only significant modification to the structure is the terminal coil being pulled tight rather than loose as in stage I. At stage III the structure is relaxed after much of the tension has been relieved. At this point, the two β sheet domains have slid relative to each other and are no longer strongly connected. Stage IV shows a peeled structure where the β sheet region has been split in two and pulled to either side of the protein. Stage V shows a further stretched structure, with high tension in the remaining structured region. At stage VI, only a couple of secondary structures remain, which are finally completely stretched at stage VII.

The peak was selected by our script, which is made available at https://kiharalab.org/SMD/. For automated peak selection, we first smoothed the data using a moving average with a window sise of 10, the equivalent of 10ps. Using this smoothed data, we then detected the force peak *F(t)* at time *t* that satisfies the following criteria:

$$t > 10ps$$

$$AND$$

$$F(t) > F(t-1ps) > F(t-2ps) > F(t-3ps) > F(t-4ps) > F(t-5ps)$$

$$AND$$

$$F(t) > F(t+1ps) > F(t+2ps) > F(t+3ps) > F(t+4ps) > F(t+5ps)$$

Thus, the force at that time step was greater than its neighbors of within 5 ps and that appeared after 10 ps.

**Logistic regression to predict GDT-TS**

We also predicted GDT-TS of models from break forces observed in the native, all-alanine, and reversed-sequence models of the same conformations using logistic regression. For prediction of models in a target, all the models from other targets were used. Parameters of the regression models and the mean square error of the predictions of the target proteins are provided in Supplementary Table 2.

**Results**

**Examples of the break force relative to the model quality**

The main idea of this work is to use the break force from steered MD to select high quality protein structure models. In Figure 2, we show examples of structure models with varying quality as defined by GDT-TS. Panel a shows the force profile of four models of T0659, a 89 residue-long single domain protein of a β-sandwich fold. The highest quality model among them, the one with GDT-TS of 94.59, clearly has the highest peak. The other three models, with a GDT-TS of 79.73, 62.84, and 28.72, (panel c) have lower break force than the highest quality model. The second highest break force was observed for the model with 28.72 GDT-TS. Thus, interestingly, the break force does not have a clear overall correlation to GDT-TS of models.

Panel b shows break force distribution of all 253 models of this target. It can be seen that models with a high break force have high GDT-TS, around 90. Thus, if models are sorted by their break force, top ranked models all have a GDT-TS around 90. However, there are high quality models with lower break forces, e.g. below 1500 pN, the same level of break force as models of 30 GDT-TS. Later we will discuss differences in high-quality models with relatively high and low break forces.

**Predictive Capabilities**

The underlying question of this work was if the physical stress given by pulling the protein chain can be used to determine relative model quality of computational models. To answer the question, for each target (Table 1), we selected the five models with the highest break force magnitudes and counted the number of high-quality models with a GDT-TS score of 80 or higher among them (Figure 3, Supplementary Table 1). Also shown are the results from the same analysis with all alanine sequence models and reversed sequence models. We evaluated the model selection performance by choosing five models following the CASP structure prediction evaluation, where participants are asked to submit five models for a target protein. As reference, the selection by break force was compared with random selection of five models. The random selection was performed 1000 times, and the average counts from the 1000 selections were reported.

Figure 3a compares the model selection results using break force values with random selection. Out of 24 targets, using break force showed better performance than random for 21 targets (87.5%). Among the 21 targets, in 19 cases all five selections, thus, 100%, had over 80 GDT-TS. In Figure 3b and 3c, we compared the selection on the native sequence models with all-alanine models and reversed sequence-models, respectively. For both cases, the model selection worked better for the native sequence models, indicating that specific amino acid interactions local structures in the native sequence models contributed to stabilize the near-native structures relative to other models. In Figure 3d, we compared the model selection from the native sequence models and the logistic regression models, which combines the break force measured on the native sequence model, the all-alanine model, and the reversed sequence-model. The regression models tied on 16 models and underperformed against the native sequence model force selection on the remaining eight. This underperformance by the regression model appears to be a result of at least two factors; the high performance of the native alone leaves little room for improvement by the regression model in the 80 and over GDT-TS category and also a larger spread of highly stable models across lower break-force models in the all-alanine model simulation (data not shown).

Comparison between using all-alanine and reversed-sequence models against random selection is provided in Table 2. As shown, using these two models worked better than random for the majority of the targets. These results imply that near-native conformation have features that can increase break force even for all-alanine or reversed sequences. The most frequently observed characteristic of highly stable reverse sequence models and all alanine models were the presence of a closed-off hydrophobic core. It was observed that native models were often stabilized by hydrophobic cores, and even destabilized by initial models that opened such cores to solvent in otherwise near-native conformations, so when there was conservation of a hydrophobic core in reverse sequence and all alanine models similar stability would be understandable.

While the majority of native target sets showed strong enrichment, there were some exceptions (Supplementary Table 1). The target set T0815, for instance, had two models with a GDT-TS of 80 or above in the top 5 highest break force selection, while random selection produced nearly twice that. However, the mean GDT-TS of the top 5 selected for native models was higher than the mean GDT-TS of the dataset, 80.236 and 73.792 respectively, showing reasonable selectivity. Similarly, the mean GDT-TS of the other two underperforming targets in terms of top 5 selection with GDT-TS of 80 or above also had higher mean GDT-TS than the dataset: T0714 has a mean GDT-TS of 83.58 for selected native models against the dataset mean of 71.96, and T0762 has a mean GDT-TS of 82.57 for selected native models against the dataset mean of 72.85. Thus, even when native peak force elevation did not produce as many models with GDT-TS of 80 or above than random selection, the average quality of a model in the top 5 using native peak selection was higher than GDT-TS of 80.

**Physical Characteristics**

As discussed in the previous section, break force can select models with high GDT-TS. However, as shown in Figure 2b, it was noticed that models with GDT-TS of 80 or higher also include those with low break force. To determine what caused the variation of break force values among models with similar high GDT-TS, we explored physical characteristics of these high scoring models.

We considered seven features (Supplementary Table 2): First, we noticed that among many target sets there was a noticeable variation in exposed hydrophobic surfaces across models with similar GDT-TS. To quantify this, we calculated the solvent accessible surface areas of hydrophobic residues, using the Kyte Doolittle hydrophobicity index[39] as reference. For each model we then multiplied the exposed area of each hydrophobic residue of the structure after equilibration by its respective hydrophobicity index value and summed then together. We refer to the resulting value as the initial Hydrophobic Solvent Accessible Surface Area (iHSASA). The other six features are the changes in energy terms, bond length, bond angle, dihedral angle, improper, conformation, and charge-charge interaction energies, respectively, computed with the NAMDEnergy plugin in VMD[38], before and after the break force peak (see Supplementary Table 2).

We performed principal component analysis (PCA) using the data collected in Supplementary Table 2, which are Pearson correlations between the break force values and the changes in energy terms and iHSASA (Figure 4a). Among the seven features considered in PCA, we colored the data points with three features that apparently correlated with PC1 or PC2 axes. In Figure 4b, data points (targets) are colored by the correlation with the bonded energy, which considers bond lengths, bond angles, dihedrals, and a peptide bond angle energy ("improper" in NAMD). Figure 4b shows correlation against the self-electric interaction energy (i.e. interactions within a chain, not protein-water interactions) and Figure 4c shows the correlation with iHASA.

High correlation with the bonded energy may be observed when a significant rearrangement of ordered structures occurs near the time of their respective break force (Figure 5). In Figure 5a, the change in bonded energy and the break force are plotted for high quality models (GDT-TS $\geq$ 80) in the target set T0798 as an example of a target set with high correlation (0.4271; Supplementary Table 2). Panel b and c in Figure 5 illustrate two models of T0798, one with a low break force (Fig. 5b, T0798TS117_1) and another one with a high break force (Fig. 5c, T0798TS430_1), but have similarly high GDT-TS (92.15 and 91.86, respectively. Panel d and e show a snapshot of unfolding trajectories of these two models. Corresponding regions in these four structures are colored in the same way to clarify the structural differences.

For the first model (T0798TS117_1) that had a lower break force, the deformation of a β sheet (pink) occurred smoothly coupled with gentle unfolding around a hinge-like coil domain (yellow). In contrast, the second model with a strong break force (T0798TS430_1) has a more compact conformation with a helical region (red) and the loop region (green) than the weaker model, which caused more interactions between them. During the unfolding process, this high degree of interaction between the helix and the loop appeared to prevent movement of the hinge (yellow), forcing unfolding through a more energetically steep pathway. In the crystal structures of PDB: 4ojk (the reference structure for this CASP model), 2f9m, and 2f9n, this helix and loop region are in the binding pocket for GTP.

In Figure 6a, the change in electrostatic energy and break force are plotted for high quality models (GDT-TS $\geq$ 80) in the target set T0773 as an example of a target set with high correlation (0.4587; Supplementary Table 2). For this target, a variation of the relative proximity of acidic and basic residues was observed across models and those with high break force magnitudes tended to have these charged residues pull away from one another at or shortly after the time of the break force peak (Figure 6).

Panels b and c in Figure 6 show two models for T0773, one with a low break force (Figure 6b, T0773TS457_4) and one with a high break force (Figure 6c, T0773TS479_1). Both models in Figure 6b and c have similarly high GDT-TS (86.36 and 80.52, respectively). Three conformations of Glu13 (shown in red, yellow, and orange) and Lys58 (shown in blue, violet, and magenta) in the two models, which were observed to have very different interactions along the unfolding pathway around the time of break force are shown in different colors. For the weaker model (T0773TS457_4; Fig. 6b), these two residues were distant in the initial conformation and did not move and thus did not contribute much to the break force. In contrast, the stronger model (T0773TS479_1; Fig. 6c), which starts with these residues nearer (~4.7-4.9 Å) to one another showed a strong interaction, where in the structure hit its breaking tension and underwent significant rearrangement only as and after these residues, Lys58 and Glu13, pulled away from one another. In the native structure of this target determined by NMR (PDB ID: 2n2u) Lys58 and Glu13 are at an average distance of 3.68 Å from one other over the NMR frames.

Figure 7 shows strong inverse correlation between iHSASA, which accounts for protein-solvent interaction. In Figure 7a, the initial hydrophobic accessible surface area (iHSASA), and break force are plotted for high quality models (GDT-TS ≥ 80) in the target set T0644 as an example of a target set with high, inverse, correlation (-0.731, Supplementary Table 2). Panels b and c in Figure 7 show two models of T0644, one with a low break force (Figure 7b, T0644TS079_5) and one with a high break force (Figure 7c, T0644TS405_5), both with high GDT-TS (82.62 and 82.45, respectively). For the first model (T0644TS079_5) with a lower break force, the iHSASA was 8975.3 $Å^2$, due in most part to the exposed hydrophobic side of its terminal helix shown on the front of the structure. In contract, Figure 7c shows the model T0644TS405_5, which had nearly double the break force, and an initial iHSASA of 5826.2 $Å^2$. The terminal helix in Figure 7c is tucked onto the bottom of the structure, and its hydrophobic side is buried against a hydrophobic patch on the rest of the structure that was exposed on T0644TS079_5. The movement of this terminal residue in T0644TS405_5 corresponds in time with the break force peak of the trajectory, and results in a dramatic spike in hydrophobic solvent accessible surface area.

**Discussion**

In this work we have demonstrated that by stressing models in molecular dynamics simulations, high quality models can be detected based on their stability. Models which required the most force to rearrange tended to have near-native conformations, though not all near-native conformations required large amounts of force to rearrange.

In addition, we have characterized some of the underlying mechanisms determining the stability of these models. First, large changes in bonded energy were observed to occur in models which had some interaction short-circuiting the tension and preventing unfolding. A large spike in bonded energy may be alleviated by changing the direction of pulling. Electrostatic energy of the protein interacting with itself was observed in models where basic and acid sidechains were near one another at the start of the production pulling runs and did not pull apart until break force peak occurred, which models with similar GDT-TS but lower break forces started with the same sidechains further apart. SASA of hydrophobic residues inversely correlated with stability. This stabilizing mechanism would come from forcing rearrangement of both protein and water molecules in such a way that water would end up in an unfavorable hydrophobic environment. However, models that already had water in these environments, i.e. in contact with hydrophobic patches on the protein, would not have this kind of environmental change to go through so would not be so resistant to rearrangement.

Kingsley et al.[40] showed that the computational pulling procedure is also useful to indicate the quality of pairwise protein-protein docking models for a few cases. It may be also extended to evaluate models of protein complexes with more subunits. The computational pulling procedure, together with other computational methods to stress protein structures, such as by applying pressure[41,42] would be also valuable in protein design and protein engineering, where developing new proteins or protein complexes with desired stability may be an important and interesting task.

**Acknowledgements**

The authors would like to acknowledge Amitava Roy for consulting on the molecular dynamics simulation setup and interpretation. This work was partly supported by the National Institutes of Health (R01GM133840, R01GM123055, and 3R01GM133840-02S1) and the National Science Foundation (CMMI1825941, MCB1925643, and DBI2003635). LM is supported by Bilsland Dissertation Fellowship from the Department of Biological Sciences, Purdue University.

**Table 1.** The dataset of protein structure models.

| Target | PDB | Length | Total | 100-80 | 80-60 | 60-40 | 40-20 | >20 |
|---|---|---|---|---|---|---|---|---|
| T0644 | 4fr9 | 158 | 549 | 156 (28.4) | 208 (37.9) | 24 (4.4) | 118 (21.5) | 43 (7.8) |
| T0650 | 5fmz | 333 | 241 | 103 (42.7) | 71 (29.5) | 40 (16.6) | 17 (7.1) | 10 (4.1) |
| T0659 | 4esn | 89 | 253 | 193 (76.3) | 10 (4.0) | 6 (2.4) | 42 (16.6) | 2 (0.8) |
| T0689 | 4fvs | 225 | 240 | 158 (65.8) | 23 (9.6) | 1 (0.4) | 28 (11.7) | 30 (12.5) |
| T0700 | 4hfx | 76 | 585 | 106 (18.1) | 174 (29.7) | 288 (49.2) | 17 (2.9) | 0 (0.0) |
| T0709 | 6mm4 | 33 | 552 | 439 (79.5) | 41 (7.4) | 64 (11.6) | 8 (1.4) | 0 (0.0) |
| T0711 | 2m7t | 33 | 565 | 294 (52.0) | 146 (25.8) | 79 (14.0) | 46 (8.1) | 0 (0.0) |
| T0712 | 4gbs | 203 | 248 | 162 (65.3) | 26 (10.5) | 16 (6.5) | 0 (0.0) | 44 (17.7) |
| T0714 | 2lvc | 88 | 271 | 165 (60.9) | 82 (30.3) | 18 (6.6) | 5 (1.8) | 1 (0.4) |
| T0716 | 2ly9 | 70 | 276 | 237 (85.9) | 26 (9.4) | 13 (4.7) | 0 (0.0) | 0 (0.0) |
| T0731 | 2lz1 | 78 | 259 | 196 (75.7) | 30 (11.6) | 19 (7.3) | 11 (4.2) | 3 (1.2) |
| T0738 | 4is2 | 239 | 264 | 195 (73.9) | 48 (18.2) | 1 (0.4) | 19 (7.2) | 1 (0.4) |
| T0749 | 4gl3 | 429 | 254 | 194 (76.4) | 8 (3.1) | 0 (0.0) | 16 (6.3) | 36 (14.2) |
| T0757 | 4gak | 243 | 236 | 31 (13.1) | 165 (69.9) | 22 (9.3) | 10 (4.2) | 8 (3.4) |
| T0762 | 4q5t | 272 | 178 | 121 (68.0) | 46 (25.8) | 8 (4.5) | 1 (0.6) | 2 (1.1) |
| T0766 | 4q52 | 127 | 181 | 128 (70.7) | 34 (18.8) | 16 (8.8) | 0 (0.0) | 3 (1.7) |
| T0773 | 2n2u | 66 | 544 | 104 (19.1) | 197 (36.2) | 132 (24.3) | 107 (19.7) | 4 (0.7) |
| T0797 | 4ojk | 32 | 469 | 278 (59.3) | 133 (28.4) | 45 (9.6) | 13 (2.8) | 0 (0.0) |
| T0798 | 4ojk | 189 | 422 | 386 (91.5) | 9 (2.1) | 8 (1.9) | 10 (2.4) | 9 (2.1) |
| T0801 | 4piw | 367 | 188 | 129 (68.6) | 42 (22.3) | 10 (5.3) | 5 (2.7) | 2 (1.1) |
| T0811 | - [a] | 180 | 183 | 158 (86.3) | 18 (9.8) | 0 (0.0) | 5 (2.7) | 2 (1.1) |

| T0815 | 4u13 | 106 | 177 | 132 (74.6) | 42 (23.7) | 0 (0.0) | 1 (0.6) | 2 (1.1) |
| T0861 | 5j5v | 324 | 181 | 143 (79.0) | 15 (8.3) | 3 (1.7) | 8 (4.4) | 12 (6.6) |
| T0891 | 4ymp | 125 | 172 | 124 (72.1) | 24 (14.0) | 8 (4.7) | 4 (2.3) | 12 (7.0) |

The model sets were taken from CASP 10-12. The number of models was counted for five GDT-TS ranges, 100 to 80, 80-60, 60-40, 40-20, and less than 20. The percentage of the total is shown in parentheses. a), The PDB entry ID is not available for this target. GDT-TS values of all the models from all the targets were taken from the data file that associated with the model structure files downloaded from the CASP website.

**Table 2. Comparison of model selection performance with random selection.**

|  | Better | Worse |
|---|---|---|
| Native | 21 | 3 |
| All-Alanine | 16 | 8 |
| Reversed-Sequence | 15 | 9 |
| Logistic Regression | 20 | 4 |

The number of target sets where each model performed better or worse in selecting high quality models of 80 or larger GDT-TS among the top five selections are listed. There are 24 targets in total.

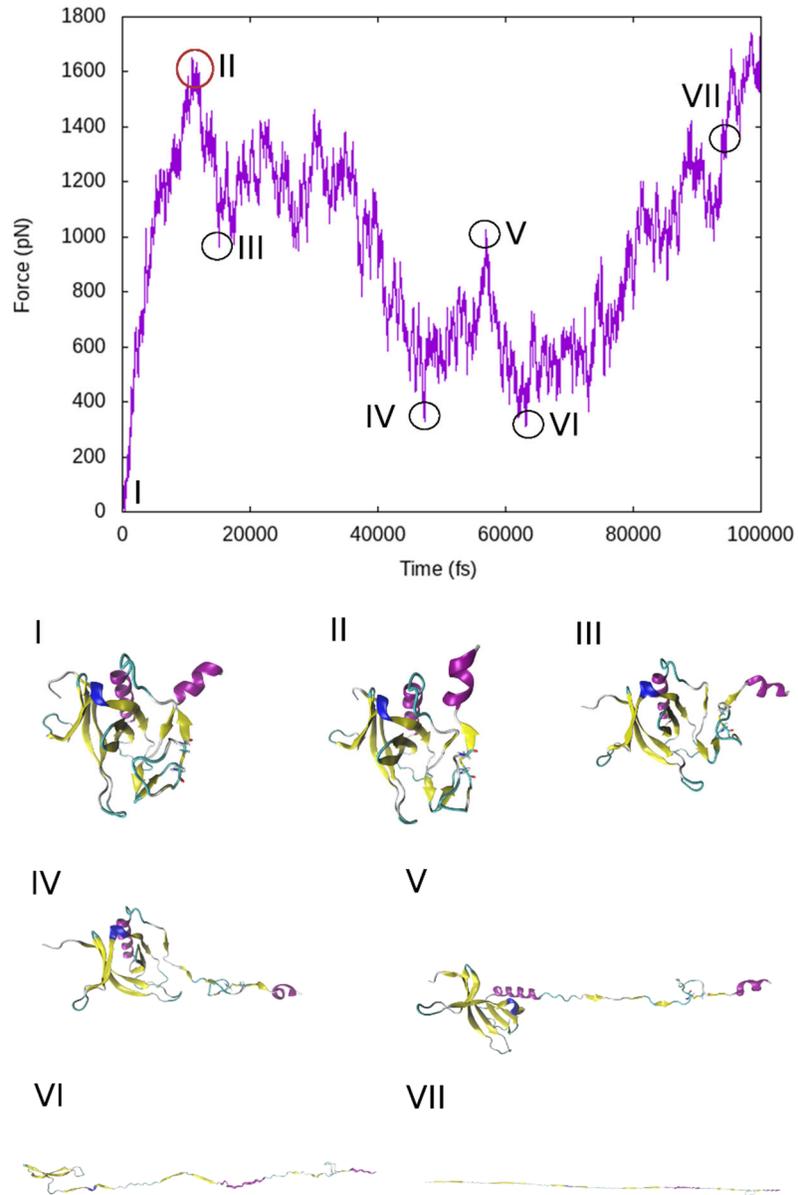

**Figure 1**. Force overtime with snap shots of the structure model T0659TS439_2 through its SMD pulling trajectory. Corresponding conformation of the model at their respective points in the trajectory are shown. Large conformational shifts are seen from II to III. The peak selected here is denoted by a red circle at II.

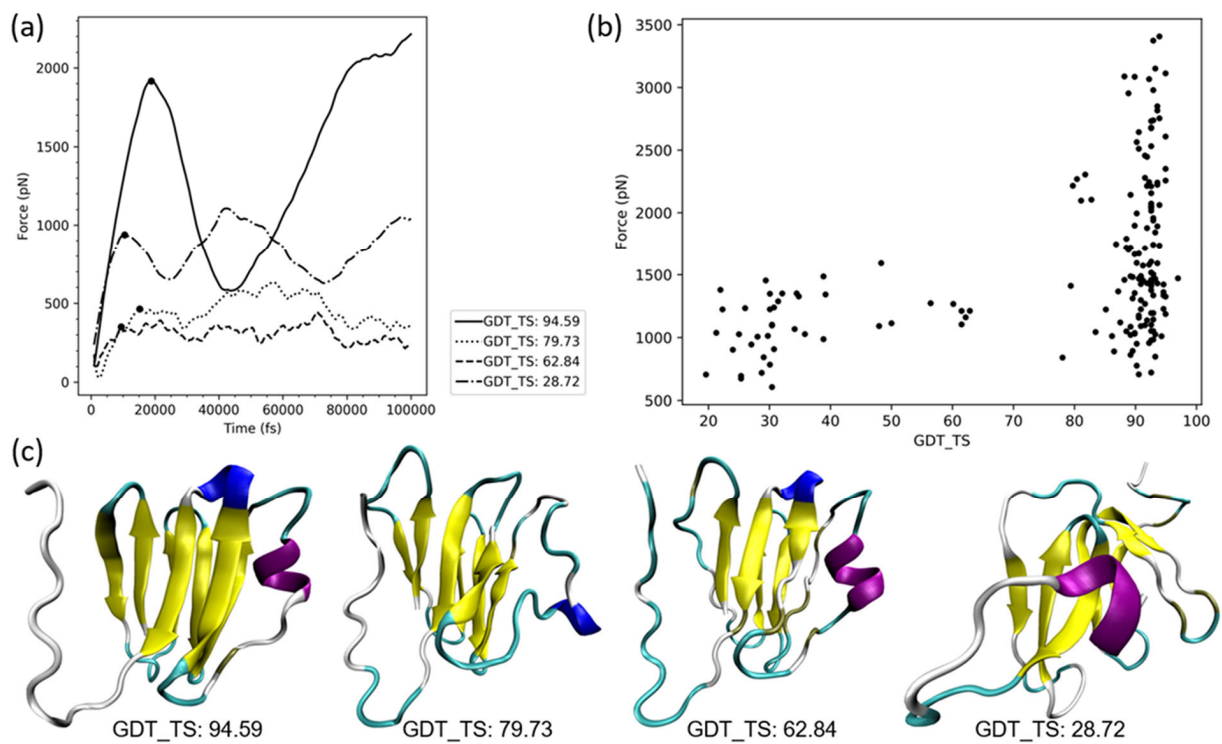

**Figure 2.** Examples of force curves of models with different GDT-TS. **(a)** Force curves of four structure models for T0659: T0659TS222_1, T0659TS035_5, T0659TS114_3, and T0659TS179_4, which had GDT-TS values of 94.69, 79.73, 62.84, and 28.72, respectively. The break force selected for a force curve is indicated with a black dot. **(b)** Peak forces of individual models of the target T0659 plotted against their starting GDT-TS. **(c)** the starting models corresponding to the curves in the panel (a) with GDT-TS scores inset.

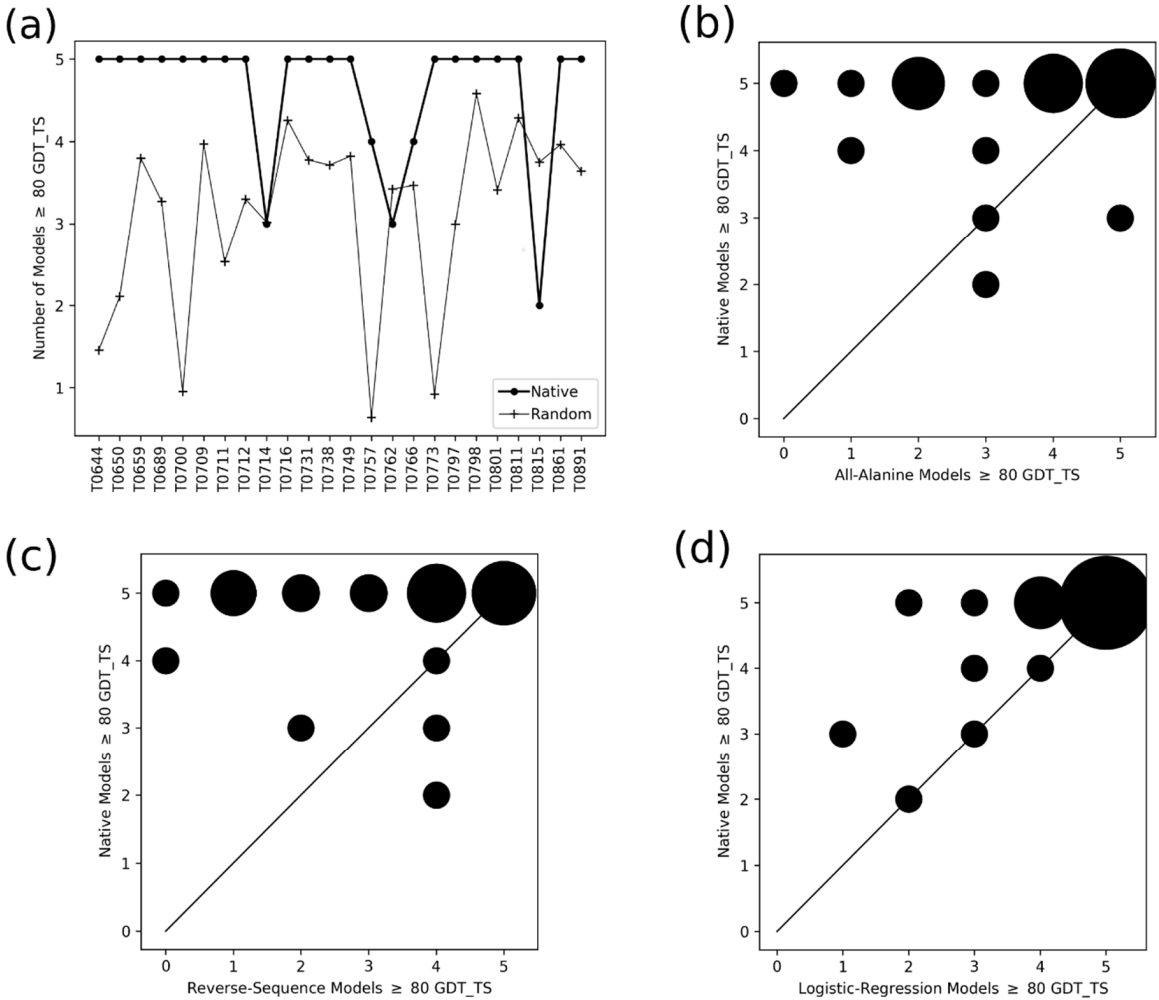

**Figure 3.** Performance of model selection with break force. For a model pool of a target protein, five models with the highest break forces were selected. **(a)** For each target, the number of high-quality models that had a GDT-TS score of 80 or higher among the top five selections were plotted. The results of the selection (black circles connected by bold lines) are compared with random selection (crosses connected by thin lines). **(b)** Comparison of the model selection performance for all-alanine models (x-axis) and the native sequence models (y-axis). For both cases, the number of high-quality models with a GDT-TS of 80 or higher among the top five selections was counted. The area of bubbles is proportional to the number of target sets at the same coordinates. **(c)** Model selection performance comparison for the reversed-sequence models (x-axis) and the native sequence models (y-axis). **(d)** Model selection performance comparison for the logistic regression (x-axis) and the native sequence models (y-axis).

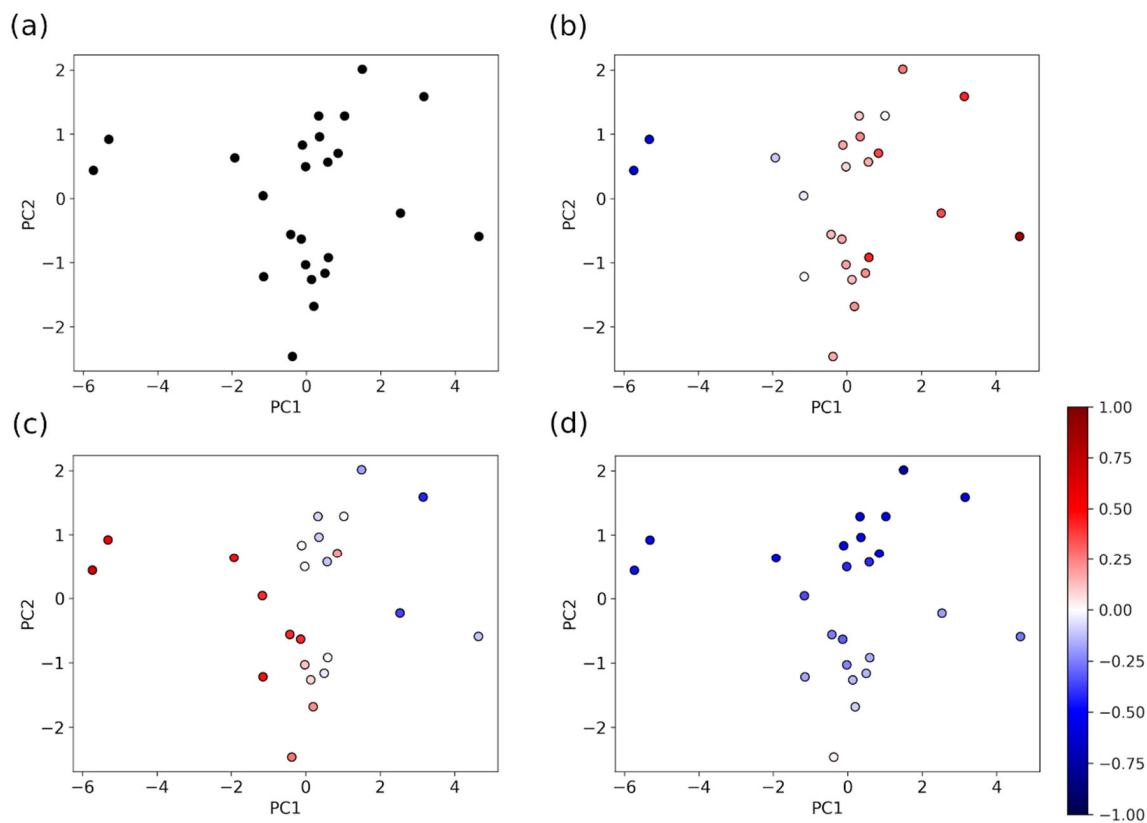

**Figure 4.** Principle component analysis of physiochemical properties of models with GDT-TS of 80 or higher. A data point corresponds to a target set. **(a)** Plot of principle component 1 (PC1) against principal component 2 (PC2) generated from using Pearson correlations of iHSASA and energy components against break force peak magnitude. **(b)** the same plot as panel a colored according to the Pearson correlation of change in the bonded energy against the magnitude of the break force peak. **(c)** the same plot colored according to the Pearson correlation of change in electrostatic energy. **(d)** the same plot colored according to the Pearson correlation of the initial hydrophobic SASA against the magnitude of the break force peaks. Color legend inset on lower

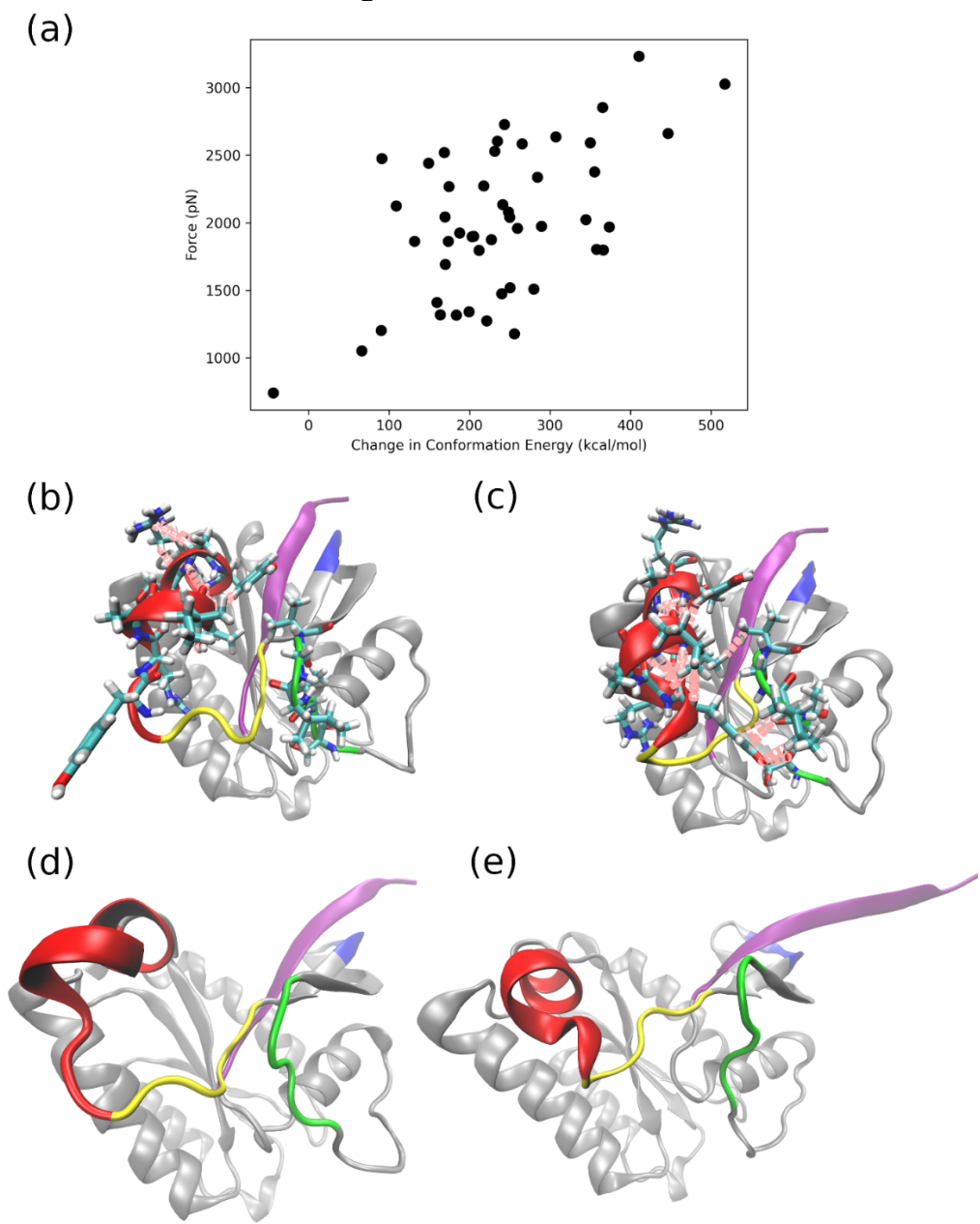

**Figure 5.** Break Force and the conformation energy. (**a**) Change in conformation energy of each model in the T0798 target set with GDT-TS of 80 or higher before and after their respective break force peak against the magnitude of the break force peak. (**b**) Structure of T0798TS117_1. A break force of 1204.220 pN and GDT-TS of 92.15, after equilibration. Residues 1 through 16 are shown in purple. Residues 52 through 56 are shown in blue Residues 66 through 75 are shown in in red with licorice representations for sidechains. Residues 37 through 39 are shown in green with a licorice representation for sidechains. Hydrogen bonds are shown in pink. The coil from residue 61 to 65 is highlighted in yellow. (**c**) Structure of T0798TS430_1, break force of 3231.310 pN and GDT-TS of 91.86, after equilibration. Color code is the same as in (b). (**d**) T0798TS117_1 at step 100 after significant unfolding. (**e**) T0798TS430_1 at step 100 after significant unfolding.

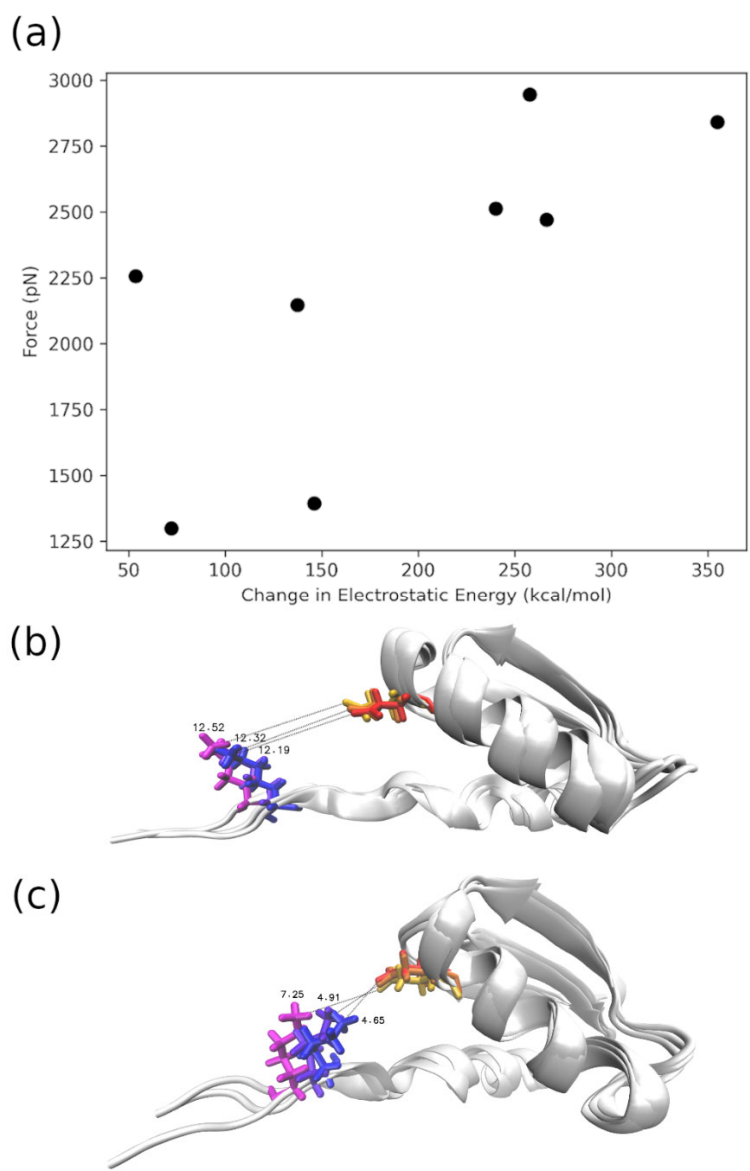

**Figure 6.** Change in electrostatic energy relative to break force. (**a**) comparison of the difference of the electrostatic energy before and after break force peak time and break force peak magnitude for the 8 model in the T0773 target set with GDT-TS of 80 or higher. (**b**) T0773TS457_4, the model with the lowest break force in the T0773 target set with a GDT-TS greater than 80. T0773TS457_4 had a break force peak of 1299.229 pN, a change in the electrostatic energy of 72.180 kcal/mol, and a GDT-TS of 86.36. Here we show three frames of the pulling trajectory, one at one picosecond before the time of the break peak, one at the time of the break peak, and one at two picoseconds after the break peak, which corresponds to the window used for the PCA above. Residue Lys58 is shown in a licorice representation and colored blue, violet, and magenta on first, second, and third frame shown, respectively. Residues Glu13 is also shown in licorice, and is colored red, orange, and yellow on the first, second, and third frame shown, respectively. The rest of the structure is shown as cartoon in white. The distance between the nitrogen in Lys58 and the nearest oxygen in Glu13 was 12.19 Å before the break peak, 12.32 Å at the break peak, and 12.52 Å after the break peak. (**c**) T0773TS479_1, the model with the highest break force in the T0773 target set with a GDT-TS greater than 80. T0773TS479_1 had a break force peak of 2945.653 pN, a change in energy of 257.710 kcal/mol, and a GDT-TS of 80.52. The frames shown relative to the break peak and representation are the same as in (b). The distance between the nitrogen in Lys58 and the nearest oxygen in Glu13 was 4.65 Å before the break peak, 4.91 Å at the break peak, and 7.25 Å after the break peak.

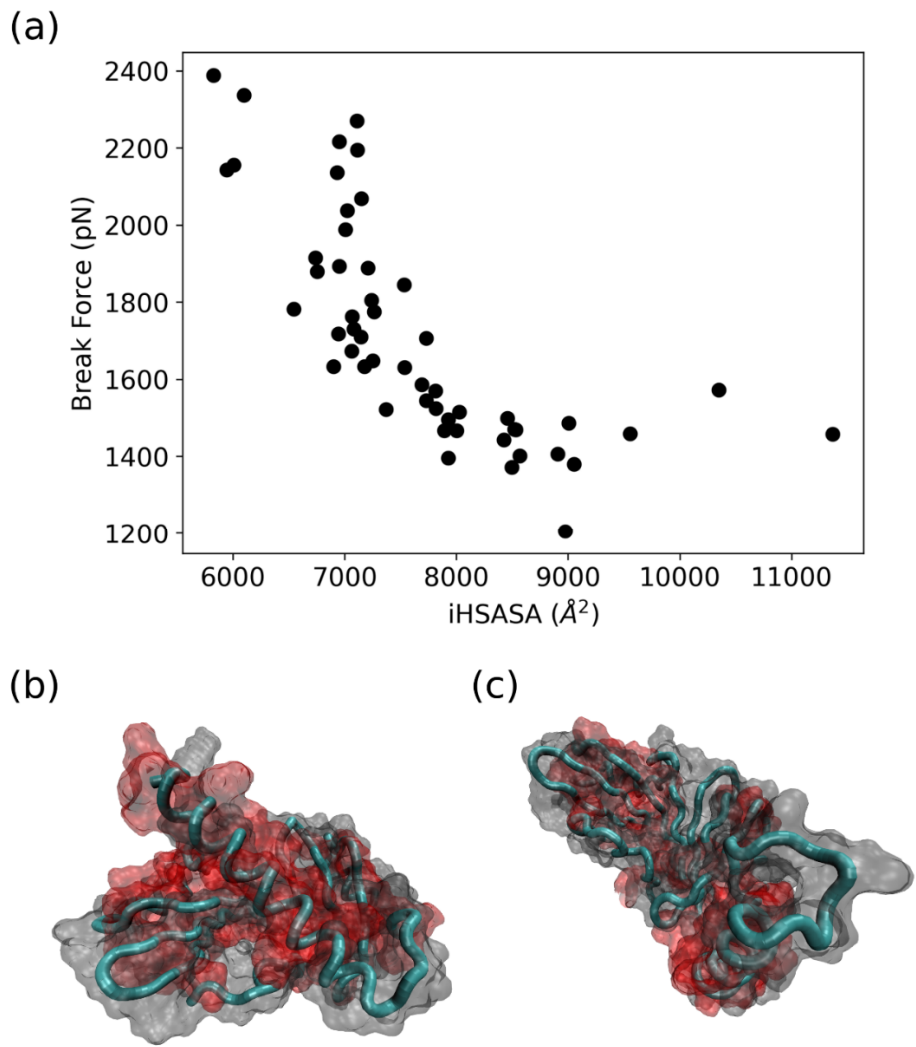

**Figure 7**. Hydrophobic solvent accessible surface area (iHSASA) and break force. Solvent accessible hydrophobic resides were calculated from the final frame of the equilibration before the pulling production run. (**a**) comparison of iHSASA and break force for all 54 models in the T0644 target set with GDT-TS of 80 or higher. (**b**) T0644TS079_5, the model with the lowest break force in the T0644 target set with a GDT-TS greater than 80. T0644TS079_5 had a break force peak of 1205.138 pN, an iHSASA of 8975.3 Å$^2$, and a GDT-TS of 82.62. The backbone is shown in cyan, and hydrophobic surfaces shown in red. (**c**) T0644TS405_4, the model with the highest break force in the T0644 target set with a GDT-TS greater than 80. T0644TS405_5 had a break force peak of 2388.648 pN, an iHSASA of 5826.2 Å$^2$, and a GDT-TS of 82.45. structures in panel b and c are aligned by all residues excluding those in the terminal helix which in a different position in the two models. In panel b, the helix is on the face of the structure while in panel b it if folded to the right, which made the appearance of the overall surfaces of the two models.